

\catcode`\@=11
\expandafter\ifx\csname @iasmacros\endcsname\relax
	\global\let\@iasmacros=\par
\else	\endinput
\fi
\catcode`\@=12


\def\rmb{\seventeenrm}

\def\singlespace{\baselineskip=\normalbaselineskip}
\def\halfspace{\baselineskip=1.5\normalbaselineskip}
\def\doublespace{\baselineskip=2\normalbaselineskip}


\def\AB{\bigskip\parindent=40pt
        \centerline{\bf ABSTRACT}\medskip\halfspace\narrower}
\def\AE{\bigskip\nonarrower\doublespace}
\def\nonarrower{\advance\leftskip by-\parindent
	\advance\rightskip by-\parindent}


\def\undertext#1{$\underline{\smash{\hbox{#1}}}$}
\def\boxit#1{\vbox{\hrule\hbox{\vrule\kern3pt
	\vbox{\kern3pt#1\kern3pt}\kern3pt\vrule}\hrule}}

\def\hence{\leavevmode\hbox{\bf .\raise5.5pt\hbox{.}.} }

\def\dalemb#1#2{{\vbox{\hrule height.#2pt
	\hbox{\vrule width.#2pt height#1pt \kern#1pt \vrule width.#2pt}
	\hrule height.#2pt}}}
\def\gtorder{\mathrel{\raise.3ex\hbox{$>$}\mkern-14mu
             \lower0.6ex\hbox{$\sim$}}}
\def\ltorder{\mathrel{\raise.3ex\hbox{$<$}\mkern-14mu
             \lower0.6ex\hbox{$\sim$}}}

\newdimen\fullhsize
\newbox\leftcolumn
\def\twoup{\hoffset=-.5in \voffset=-.25in
  \hsize=4.75in \fullhsize=10in \vsize=6.9in
  \def\fullline{\hbox to\fullhsize}
  \let\lr=L
  \output={\if L\lr
        \global\setbox\leftcolumn=\columnbox\global\let\lr=R \advancepageno
      \else \doubleformat \global\let\lr=L\fi
    \ifnum\outputpenalty>-20000 \else\dosupereject\fi}
  \def\doubleformat{\shipout\vbox{
    \fullline{\box\leftcolumn\hfil\columnbox}\advancepageno}}
  \def\columnbox{\leftline{\vbox{\makeheadline\pagebody\makefootline}}}
  \tolerance=1000 }
\catcode`\@=11					



\font\fiverm=cmr5				
\font\fivemi=cmmi5				
\font\fivesy=cmsy5				
\font\fivebf=cmbx5				

\skewchar\fivemi='177
\skewchar\fivesy='60


\font\sixrm=cmr6				
\font\sixi=cmmi6				
\font\sixsy=cmsy6				
\font\sixbf=cmbx6				

\skewchar\sixi='177
\skewchar\sixsy='60


\font\sevenrm=cmr7				
\font\seveni=cmmi7				
\font\sevensy=cmsy7				
\font\sevenit=cmti7				
\font\sevenbf=cmbx7				

\skewchar\seveni='177
\skewchar\sevensy='60


\font\eightrm=cmr8				
\font\eighti=cmmi8				
\font\eightsy=cmsy8				
\font\eightit=cmti8				
\font\eightbf=cmbx8				

\skewchar\eighti='177
\skewchar\eightsy='60


\font\ninei=cmmi9
\font\ninesy=cmsy9

\skewchar\ninei='177
\skewchar\ninesy='60


\font\tenrm=cmr10				
\font\teni=cmmi10				
\font\tensy=cmsy10				
\font\tenex=cmex10				
\font\tenit=cmti10				
\font\tensl=cmsl10				
\font\tenbf=cmbx10				
\font\tentt=cmtt10				
\font\tenss=cmss10				
\font\tensc=cmcsc10				
\font\tenbi=cmmib10				

\skewchar\teni='177
\skewchar\tenbi='177
\skewchar\tensy='60

\def\tenpoint{\ifmmode\err@badsizechange\else
	\textfont0=\tenrm \scriptfont0=\sevenrm \scriptscriptfont0=\fiverm
	\textfont1=\teni  \scriptfont1=\seveni  \scriptscriptfont1=\fivemi
	\textfont2=\tensy \scriptfont2=\sevensy \scriptscriptfont2=\fivesy
	\textfont3=\tenex \scriptfont3=\tenex   \scriptscriptfont3=\tenex
	\textfont4=\tenit \scriptfont4=\sevenit \scriptscriptfont4=\sevenit
	\textfont5=\tensl
	\textfont6=\tenbf \scriptfont6=\sevenbf \scriptscriptfont6=\fivebf
	\textfont7=\tentt
	\textfont8=\tenbi \scriptfont8=\seveni  \scriptscriptfont8=\fivemi
	\def\rm{\tenrm\fam=0 }%
	\def\it{\tenit\fam=4 }%
	\def\sl{\tensl\fam=5 }%
	\def\bf{\tenbf\fam=6 }%
	\def\tt{\tentt\fam=7 }%
	\def\ss{\tenss}%
	\def\sc{\tensc}%
	\def\bmit{\fam=8 }%
	\rm\setparameters\setbaselines\fi}


\font\twelverm=cmr12				
\font\twelvei=cmmi12				
\font\twelvesy=cmsy10	scaled\magstep1		
\font\twelveex=cmex10	scaled\magstep1		
\font\twelveit=cmti12				
\font\twelvesl=cmsl12				
\font\twelvebf=cmbx12				
\font\twelvett=cmtt12				
\font\twelvess=cmss12				
\font\twelvesc=cmcsc10	scaled\magstep1		
\font\twelvebi=cmmib10	scaled\magstep1		

\skewchar\twelvei='177
\skewchar\twelvebi='177
\skewchar\twelvesy='60

\def\twelvepoint{\ifmmode\err@badsizechange\else
	\textfont0=\twelverm \scriptfont0=\eightrm \scriptscriptfont0=\sixrm
	\textfont1=\twelvei  \scriptfont1=\eighti  \scriptscriptfont1=\sixi
	\textfont2=\twelvesy \scriptfont2=\eightsy \scriptscriptfont2=\sixsy
	\textfont3=\twelveex \scriptfont3=\tenex   \scriptscriptfont3=\tenex
	\textfont4=\twelveit \scriptfont4=\eightit \scriptscriptfont4=\sevenit
	\textfont5=\twelvesl
	\textfont6=\twelvebf \scriptfont6=\eightbf \scriptscriptfont6=\sixbf
	\textfont7=\twelvett
	\textfont8=\twelvebi \scriptfont8=\eighti  \scriptscriptfont8=\sixi
	\def\rm{\twelverm\fam=0 }%
	\def\it{\twelveit\fam=4 }%
	\def\sl{\twelvesl\fam=5 }%
	\def\bf{\twelvebf\fam=6 }%
	\def\tt{\twelvett\fam=7 }%
	\def\ss{\twelvess}%
	\def\sc{\twelvesc}%
	\def\bmit{\fam=8 }%
	\rm\setparameters\setbaselines\fi}


\font\fourteenrm=cmr10	scaled\magstep2		
\font\fourteeni=cmmi10	scaled\magstep2		
\font\fourteensy=cmsy10	scaled\magstep2		
\font\fourteenex=cmex10	scaled\magstep2		
\font\fourteenit=cmti10	scaled\magstep2		
\font\fourteensl=cmsl10	scaled\magstep2		
\font\fourteenbf=cmbx10	scaled\magstep2		
\font\fourteentt=cmtt10	scaled\magstep2		
\font\fourteenss=cmss10	scaled\magstep2		
\font\fourteensc=cmcsc10 scaled\magstep2	
\font\fourteenbi=cmmib10 scaled\magstep2	

\skewchar\fourteeni='177
\skewchar\fourteenbi='177
\skewchar\fourteensy='60

\def\fourteenpoint{\ifmmode\err@badsizechange\else
	\textfont0=\fourteenrm \scriptfont0=\tenrm \scriptscriptfont0=\sevenrm
	\textfont1=\fourteeni  \scriptfont1=\teni  \scriptscriptfont1=\seveni
	\textfont2=\fourteensy \scriptfont2=\tensy \scriptscriptfont2=\sevensy
	\textfont3=\fourteenex \scriptfont3=\tenex \scriptscriptfont3=\tenex
	\textfont4=\fourteenit \scriptfont4=\tenit \scriptscriptfont4=\sevenit
	\textfont5=\fourteensl
	\textfont6=\fourteenbf \scriptfont6=\tenbf \scriptscriptfont6=\sevenbf
	\textfont7=\fourteentt
	\textfont8=\fourteenbi \scriptfont8=\tenbi \scriptscriptfont8=\seveni
	\def\rm{\fourteenrm\fam=0 }%
	\def\it{\fourteenit\fam=4 }%
	\def\sl{\fourteensl\fam=5 }%
	\def\bf{\fourteenbf\fam=6 }%
	\def\tt{\fourteentt\fam=7}%
	\def\ss{\fourteenss}%
	\def\sc{\fourteensc}%
	\def\bmit{\fam=8 }%
	\rm\setparameters\setbaselines\fi}


\font\seventeenrm=cmr10 scaled\magstep3		


\newdimen\rp@
\newcount\@basestretchnum
\newskip\@baseskip
\newskip\headskip
\newskip\footskip


\def\setparameters{\rp@=.1em
	\headskip=24\rp@
	\footskip=\headskip
	\delimitershortfall=5\rp@
	\nulldelimiterspace=1.2\rp@
	\scriptspace=0.5\rp@
	\abovedisplayskip=10\rp@ plus3\rp@ minus5\rp@
	\belowdisplayskip=10\rp@ plus3\rp@ minus5\rp@
	\abovedisplayshortskip=5\rp@ plus2\rp@ minus4\rp@
	\belowdisplayshortskip=10\rp@ plus3\rp@ minus5\rp@
	\normallineskip=\rp@
	\lineskip=\normallineskip
	\normallineskiplimit=0pt
	\lineskiplimit=\normallineskiplimit
	\jot=3\rp@
	\setbox0=\hbox{\the\textfont3 B}\p@renwd=\wd0
	\skip\footins=12\rp@ plus3\rp@ minus3\rp@
	\skip\topins=0pt plus0pt minus0pt}


\def\setbaselines{\maxdepth=4\rp@\baselinestretch=\@basestretchnum}


\def\baselinestretch{\afterassignment\@basestretch\@basestretchnum}
\def\@basestretch{%
	\@baseskip=12\rp@ \divide\@baseskip by1000
	\normalbaselineskip=\@basestretchnum\@baseskip
	\baselineskip=\normalbaselineskip
	\bigskipamount=\the\baselineskip
		plus.25\baselineskip minus.25\baselineskip
	\medskipamount=.5\baselineskip
		plus.125\baselineskip minus.125\baselineskip
	\smallskipamount=.25\baselineskip
		plus.0625\baselineskip minus.0625\baselineskip
	\setbox\strutbox=\hbox{\vrule height.708\baselineskip
		depth.292\baselineskip width0pt }}



\def\makeheadline{\vbox to0pt{\baselinestretch=1000
	\vskip-\headskip \vskip1.5pt
	\line{\vbox to\ht\strutbox{}\the\headline}\vss}\nointerlineskip}

\def\makefootline{\baselineskip=\footskip\line{\the\footline}}

\def\big#1{{\hbox{$\left#1\vbox to8.5\rp@ {}\right.\n@space$}}}
\def\Big#1{{\hbox{$\left#1\vbox to11.5\rp@ {}\right.\n@space$}}}
\def\bigg#1{{\hbox{$\left#1\vbox to14.5\rp@ {}\right.\n@space$}}}
\def\Bigg#1{{\hbox{$\left#1\vbox to17.5\rp@ {}\right.\n@space$}}}


\mathchardef\alpha="710B
\mathchardef\beta="710C
\mathchardef\gamma="710D
\mathchardef\delta="710E
\mathchardef\epsilon="710F
\mathchardef\zeta="7110
\mathchardef\eta="7111
\mathchardef\theta="7112
\mathchardef\iota="7113
\mathchardef\kappa="7114
\mathchardef\lambda="7115
\mathchardef\mu="7116
\mathchardef\nu="7117
\mathchardef\xi="7118
\mathchardef\pi="7119
\mathchardef\rho="711A
\mathchardef\sigma="711B
\mathchardef\tau="711C
\mathchardef\upsilon="711D
\mathchardef\phi="711E
\mathchardef\chi="711F
\mathchardef\psi="7120
\mathchardef\omega="7121
\mathchardef\varepsilon="7122
\mathchardef\vartheta="7123
\mathchardef\varpi="7124
\mathchardef\varrho="7125
\mathchardef\varsigma="7126
\mathchardef\varphi="7127
\mathchardef\imath="717B
\mathchardef\jmath="717C
\mathchardef\ell="7160
\mathchardef\wp="717D
\mathchardef\partial="7140
\mathchardef\flat="715B
\mathchardef\natural="715C
\mathchardef\sharp="715D


\def\err@badsizechange{%
	\immediate\write16{--> Size change not allowed in math mode, ignored}}

\baselinestretch=1000
\tenpoint

\catcode`\@=12					
\twelvepoint
\vsize=8.5in
\hsize=6in

\overfullrule=0pt
\rightline{IASSNS-HEP-95/26}
\rightline{hep-ph/9506211}
\vfill
\bigskip
\centerline{\rmb BOSE-FERMI SYMMETRY:  A CRUCIAL ELEMENT}
\bigskip
\centerline{{\rmb IN ACHIEVING UNIFICATION}\footnote{$^\dagger$}
{\singlespace Talk presented at the Int'l. Sym. on ``Bose and Twentieth Century
Physics,''
\break
\parindent=25pt
Calcutta, India, January 1994, to appear in the Proceedings.}}
\bigskip
\vfill
\bigskip
\centerline{\bf Jogesh C. Pati}
\bigskip
\bigskip
\centerline{\bf Institute for Advanced Study}
\centerline{\bf Princeton, NJ 08540 \ USA}
\centerline{\bf and}
\centerline{{\bf Department of Physics,  University of
Maryland}\footnote{$^*$}{Permanent address.  E-mail:
``pati@umdhep.umd.edu''}}
\centerline{\bf College Park, MD  \ 20742 \ USA}
\vfill
\bigskip
\AB
This talk is dedicated to honor the memory of Professor S.N. Bose.
I survey the crucial roles played by Bose-Fermi symmetry in all recent
attempts at higher unification, which include the ideas of (i) the conventional
approach to grand unification; (ii) the preonic approach; and
(iii) superstrings.
\AE
\vfill\eject
\pageno=2

{\rmb I.~~Preliminary Remarks}
\halfspace           

I regard it a special privilege to speak at this conference which
is dedicated to
celebrate the birth centenary of a great scientist
and my country-man -- Professor Saytendra Nath Bose.  Beyond doubt,
Bose's contribution [1] to physics is one of the landmarks in the
development of
quantum theory.  In one stroke, he introduced two new concepts of
lasting value:  the concept of massless particles with two states of
polarization (these are the photons with spin 1) and the concept that
their number is not conserved; that they obey a new statistics.
In Pais' words, ``The paper by Bose is the fourth
and last of the revolutionary papers of the old quantum theory (the
other three being by respectively Planck, Einstein and Bohr)'' [2].  To
this I will add that Bose's contribution, attributing the concept of a
new statistics to
photons, turned out to be an integral feature of
relativistic quantum field theory as well through a realization that
evolved during the 1930's through the 50's.  This is the famous connection
between spin and statistics [3] which asserts that particles of integer
spins (i.e, spin 0,1,2,$\ldots$) obey what is now called the
Bose-Einstein statistics while those of half-integer spins (i.e.,spin-1/2,
3/2, 5/2, $\ldots$) obey the Fermi-Dirac statistics or the exclusion
principle.  This in turn allows us to classify all known elementary
particles as either {\it bosons} or {\it fermions} which respectively
obey the Bose-Einstein and the Fermi-Dirac statistics.

In honor of Professor Bose, I will elucidate here the role of a further
concept
in recent attempts at achieving a unification of matter and its forces.
This is the concept of a symmetry that
relates fermions to bosons, which evolved some fifty years after the
birth of quantum statistics.

As a prelude, let me first say a few words about the status of the field
of particle physics as it existed before the introduction of this
new symmetry.  One central goal of
elementary particle physics has been to search for
principles which would dictate the existence of particles and their
forces.  To cite a few examples, the principle of local gauge
invariance dictates
the existence of the photon and the gluons as well as that of the
associated forces
of quantum electro- and chromodynamics.  The
principle of general coordinate invariance proposed by Einstein leads to
the familiar gravitational ``forces'' as a consequence of the curvature
of space-time and dictates the existence of spin-2 graviton.  These
principles in turn help preserve the masslessness of these ``gauge''
particles despite quantum corrections, at least in perturbation theory.
Coming now to spin-1/2 particles, although there was no such {\it a
priori} rationale, at least not until the developments in the 70's and
the 80's for their existence, one
may advance a different type of reason which is that spin-1/2 is the
{\it smallest} unit of spin, associated with an elementary particle,
that one needs to build
all higher spin-particles as composites.  Furthermore, purely from an
utilitarian point of view, spin-1/2 particles (i.e., electrons) are at
least {\it needed} since they obey the exclusion principle which is
relevant to the explanation of chemistry and in turn to the biology of
life.  Now, once spin-1/2 particles are introduced into the
lagrangian, they have the good feature that their masses remain
protected against arbitrarily large
quantum corrections through the so-called chiral symmetry which
guarantees that the quantum corrections to the masses of spin-1/2
fermions in perturbation theory either vanish or are bounded by a
logarithmic cutoff (symbolizing short-distance physics) depending upon
whether their ``bare'' mass-terms are zero or non-zero.

By contrast, no such principle, not even utilitarian arguments, existed
until the early 70's, which would either dictate the existence of
elementary spin-0 bosons or guard their masses against large quantum
corrections.  Nevertheless, there, of course, exist spin-0 particles
which are known to be relevant in particle physics.  In particular, the
pions are the carriers of the nuclear force and the Higgs bosons (yet to
be discovered) induce spontaneous breaking of the electroweak symmetry
and give masses to W and Z bosons.  As regards the pions, it is
known that they are {\it not} elementary.  They can be
identified as the (pseudo) goldstone bosons associated with a dynamical
breaking of the chiral symmetry of the up and down quarks and thus can
be viewed as $q\bar q$ composites.  For this reason, the
mass of the pion does not get large quantum corrections.  It is
controlled by the relevant chiral symmetry breaking parameter which is
determined by (a) the finite ``bare'' masses of the up and down quarks
and (b) the QCD scale-parameter.

What about the origin of the Higgs-boson?  It is not yet known
(especially if one allows for the preonic idea to be mentioned shortly)
whether the Higgs boson is
elementary or composite.  The idea of compositeness of the Higgs boson
in the sense of technicolor, which treats quarks, leptons and
technifermions as elementary but Higgs boson as composite is perhaps
excluded because it runs into difficulties with flavor-changing neutral
current processes and oblique electroweak parameters.  As I shall
elucidate later in this talk, no such difficulty exists, however, and one
obtains a viable and economical picture, if one assumes that, together
with
the Higgs boson, the quarks and the leptons are composite as well,
sharing common constituents -- called ``preons''.  Thus, either the
Higgs boson is composite in the context of a certain preonic theory,
or it is elementary.  In case it is elementary, which is in fact the
conventional view,
the two pertinent questions are:  (i) What if any is
an {\it a priori} rationale for its existence, and , equally
important, (ii) how can one protect its mass against
large quantum corrections?  To be specific, since the Higgs boson
couples to the gauge bosons and also possesses quartic self couplings,
one obtains corrections in one loop to its mass which are proportional
to $\alpha_i\Lambda^2$ ($\Lambda$ is the cutoff characterizing
short-distance physics and $\alpha$'s are coupling parameters).
Allowing for short-distance physics to include
at least gravity and possibly grand unification, one would expect
$\Lambda \sim 10^{16}-10^{19}~GeV$.  To obtain a physical mass $\ltorder
1~ TeV$, one thus needs {\it unnatural fine tuning} by some 24 orders of
magnitude (or higher) in the choice of the bare mass of the Higgs boson
to cancel the large quantum corrections.  Such a fine tuning is unnatural,
unattractive and thus unacceptable in a fundamental theory.

The problem of this unnatural fine tuning gets resolved and the
existence of spin-0 bosons derives a significance on a par with that of
spin-1/2 fermions through the
idea of a new symmetry -- commonly called ``supersymmetry'' [4] --
which transforms spin-1/2 fermions into spin-0
bosons and {\it vice versa}.  Since it is a symmetry that transforms bosons
into fermions, I will sometimes refer to it in the
course of this talk as the ``{\it Bose-Fermi Symmetry}''.

The power of supersymmetry arises because its
generator(s) $Q$ transforms a spin-0 (or spin-1) boson into a spin-1/2 fermion
and thereby changes the spin of the particle by 1/2
unit as well as its statistics.  Thus, $Q$ transforms as a
spin-1/2 fermionic operator.  This
is in contrast to the generators of the time-honored Lie algebras,
associated with the familiar symmetries such as isospin and SU(3), whose
generators
transform as Lorentz-scalars -- i.e., as spin-0 bosonic operators; and
which can thus
transform a particle of a given spin
into another of the {\it same
spin}, only.  The fermionic generators $(Q,\bar Q)$ of $N=1$
supersymmetry and the bosonic
energy-momentum operators $P_\mu$,
together, define in fact a graded Lie algebra, consisting of a combination of
commutators and anticommutators.  In particular, they satisfy:
$\{Q,\bar Q\}=2\sigma^\mu P_\mu ;~$$\{Q,Q\}=\{\bar Q,\bar Q\}=0;~$$
[P_\mu ,Q]=[P_\mu ,\bar Q]=[P_\mu ,P_\nu]=0~.$

Because of this new feature, supersymmetry brings some major benefits:

\noindent
(i)~~First, as mentioned above, supersymmetry unites
fermions and bosons as members
of a supermultiplet and thereby provides the rationale for
the existence of spin-0 matter, {\it on a par} with that of spin-1/2
matter.

\noindent
(ii)~~Second, supersymmetry permits a non-trivial marriage of
space-time (Poincar\'e) symmetries with internal symmetries in accord
with relativity.  As shown by Haag, Lopuszanski and Sohnius [5],
the graded Lie algebra associated with supersymmetry is the only
framework within which such a marriage can be achieved consistent with
relativistic quantum field theory.
For example, SU(2)-isospin symmetry
together with supersymmetry groups the spin-1/2 doublet of
$(u,d)$-quarks with the spin-0 doublet of $(\tilde u, \tilde d)$-squarks
to make a super-bidoublet
$$
\left[\matrix{u&\leftrightarrow&\tilde u\cr
\updownarrow&&\updownarrow\cr
d&\leftrightarrow&\tilde d\cr}\right]
$$
where all four members are degenerate and are related to each other by
symmetry generators.  To judge the importance of this property of
supersymmetry, it is useful to recall past attempts of this nature
which proposed to combine particles such as ($\pi$ and $\rho$) and ($N$
and $N^*$) which differ in spin only by integer units.  These attempts
failed because, as shown by Coleman and Mandula [6], with only
bosonic symmetry operators satisfying Lie algebras, they did not satisfy
the constraints of relativity.
In short, Bose-Fermi symmetry, together with the associated graded
Lie algebra, brings about a synthesis of
fundamental matter that goes well beyond that permissible within
symmetries of just bosonic operators.
\smallskip
\noindent
(iii)~~ The third major advantage of quantum field theories with
supersymmetry is that as a rule these theories are far less ultraviolet
divergent than their non-supersymmetric counterparts.  This comes about
due to cancellation between fermionic and bosonic partners in quantum
loops in SUSY theories
[7].  Because of such cancellation, the $N=4$ supersymmetric Yang-Mills
theories turn out in fact to be ultraviolet finite, and point-particle
supergravity theories, though still non-renormalizable, exhibit much
better ultraviolet behavior than non-supersymmetric Einstein gravity.
Owing to the same
cancellation in SUSY theories, the quadratic divergence in the spin-0
boson self-energy (mass)$^2$ drops out.  Instead, the self-(mass)$^2$ of
spin-0 bosons is given by $(\alpha_i/\pi)|m_B^2-m_F^2|$, where B and F
respectively denote boson and fermion SUSY partners.  This would, of
course, vanish if SUSY were exact (i.e, $m_B=m_F$).  In practice, it is
controlled by the SUSY-breaking parameter $|m_B^2-m_F^2|$.  Assuming
that this parameter is of order 1 to 10 TeV$^2$, say, we see that in
SUSY theories one no longer needs unnatural fine tuning of the Higgs
(mass)$^2$ to keep the mass of the Higgs-boson at the 1 TeV-scale.  This
feature is thus generally regarded as one of the main {\it practical
motivations} for supersymmetry and the reason for expecting that SUSY
partners should exist at the TeV-scale.
\medskip
\noindent
(iv)~~The fourth major advantage of supersymmetry -- one that would be
most needed if the idea of preonic substructure turns out to be right
(see discussions later) -- is certain novel properties of a class of strongly
interacting supersymmetric gauge theories relative to non-supersymmetric
QCD.  To cite one such feature, the Witten index theorem forbids a
dynamical breaking of supersymmetry in SUSY SU(N)-theories with massless
matter, at least in so far as one can neglect gravity [8].  As a result,
the matter-fermion condensate $\langle \bar\psi\psi\rangle$, which
breaks not only chiral symmetry but also supersymmetry, must vanish in
such theories (at least as $M_{Planck}\to\infty$).  This is in striking
contrast to the case of ordinary QCD where the chiral symmetry-breaking
quark-pair condensate $\langle\bar q q\rangle$ does in fact form with a
normal strength.  It is this inhibition in the formation of
$\langle\bar\psi\psi\rangle$ and other SUSY-breaking condensates which
has recently been utilized to build a viable and economical preon-model
with many attractive features [9,10,11,12].  These include:  (i) an
explanation of the protection of composite quark-lepton masses compared
to their scale of compositeness [10] and (ii) a natural origin of the
hierarchy of mass-scales from $M_{Pl}\to m_W\to m_e\to m_\nu$ [9].
\medskip
\noindent
(v)~~Last but not least, the greatest benefits of Bose-Fermi symmetry is
derived in the context of all attempts at higher unification, which
include the ideas of (a) the conventional approach to grand unification,
(b) the preonic approach,
(c) supergravity and, of course, (d) superstrings.
To put in
one sentence, it seems that none of these ideas would work without the aid
of supersymmetry.  To present some of these benefits of supersymmetry
in the context of higher unification,
which is the main purpose of my talk, I need to say a few words
about the puzzles in
particle physics which confront us in the context of the standard model
and the unifying ideas which have been proposed to resolve some of these
puzzles.
\bigskip
\item{\rmb II.}
{\rmb Going Beyond the Standard Model}

The standard model of particle physics (SM) has brought a good deal of
synthesis in our understanding of the basic forces of nature, especially
in comparison to its predecessors, and has turned out to be brilliantly
successful in terms of its agreement with experiments.  Yet, as
recognized for some time [13], it falls short as a fundamental theory
because it introduces some 19 parameters.  And it
does not explain (i) family replication;
(ii) the coexistence of the two kinds of matter:  quarks {\it
and} leptons; (iii) the coexistence of the electroweak {\it and} the
QCD forces with their hierarchical strengths $g_1 \ll g_2 \ll
g_3$, as observed at low energies; (iv) quantization of electric charge;
(v) inter and intrafamily mass-hierarchies - {\it i.e.},
$m_{u, d, e} \ll m_{c,s,\mu}\ll m_{t,b,\tau}$ and $m_b\ll m_t$, etc. -
reflected by ratios such as $(m_u/m_t) \sim 10^{-4},~(m_c/m_t)\sim
10^{-2}$ and $(m_b/m_t)\sim {1\over 35}$; and (vi) the origin of diverse
mass scales that span over more than 27 orders of magnitude from
$M_{Planck}$ to $m_W$ to $m_e$ to $m_\nu$, whose ratios involve
very small numbers such as$(m_W/M_{Pl})\sim
10^{-17},~(m_e/M_{Pl})\sim 10^{-22}$ and $(m_\nu/M_{Pl})< 10^{-27}$.
There are in addition the two most basic questions:  (vii) how does
gravity fit into the whole scheme, especially in the context of a good
quantum theory?, and (viii) why is the cosmological constant so small or
zero?

These issues constitute at present some of the major puzzles of particle
physics and provide motivations for contemplating new
physics beyond the standard model which should shed light on them.
The ideas which have been proposed and which do show promise to
resolve at least some of these puzzles, include the following
hypotheses:
\medskip
(1)~~{\bf Grand Unification}: ~~ The hypothesis of grand
unification, which proposes an underlying unity of the fundamental
particles and their forces [13,14,15],
appears attractive
because it explains at once (i) the quantization of electric charge, (ii) the
existence of quarks {\it and} leptons with $Q_e=-Q_p$, and (iii) the
existence of the strong, the electromagnetic and the weak forces with
$g_3\gg g_2\gg g_1$ at low energies, but $g_3=g_2=g_1$ at high energies.
These are among the puzzles listed above and grand unification
resolves all three.
{\it Therefore I believe that the central concept of
grand unification is, very likely, a step in the right direction.}  By
itself, it does not address, however, the remaining
puzzles listed above, including the issues of family replication and
origin of mass-hierarchies.
\medskip
(2)~~{\bf Supersymmetry}:~~ As mentioned before, this is the symmetry that
relates fermions to bosons[4].  As a local
symmetry, it is attractive because it implies the existence of gravity.
It has
the additional virtue that it
helps maintain a large hierarchy in mass-ratios such as
$(m_{\phi}/M_U) \sim 10^{-14}$ and $(m_{\phi}/M_{p\ell}) \sim 10^{-17}$,
without
the need for fine tuning, provided, however, such ratios are put in by
hand.
Thus it provides a technical resolution of the gauge hierarchy problem,
{\it but
by itself does not explain the origin of the large hierarchies}.
\medskip
(3)~~{\bf Compositeness}:~~  Here there are {\it two
distinct suggestions}:
\smallskip
(a)~~\undertext{Technicolor}:  The idea of technicolor [16]
proposes that the Higgs bosons are composite but quarks and leptons are
still elementary.  Despite the attractive feature of dynamical symmetry
breaking which eliminates elementary Higgs bosons and thereby the arbitrary
parameters which go with them, this idea
is excluded, at least in its simpler versions, owing to conflicts with
flavor-changing neutral current processes and oblique electroweak
corrections.  The so-called walking technicolor models may be arranged
to avoid some of these conflicts at the expense, however, of excessive
proliferation in elementary constituents.  Furthermore, as a generic
feature, none of these models seem capable of addressing any of the
basic issues listed above, including those of family replication and
fermion mass-hierarchies.  Nor do they go well with the
hypothesis of a unity of the basic forces.
\smallskip
(b)~~\undertext{ Preons}:  By contrast, the idea of preonic
compositeness which proposes that not just the Higgs bosons but also the
quarks and the leptons are composites of a {\it common} set of
constituents called ``preons'' seems much more promising.  Utilizing
supersymmetry to its advantage, the preonic approach has evolved over
the last few years to acquire a form [9,10,11,12] which is (a) far more
economical in field-content and especially in parameters than either the
technicolor or the conventional grand unification models, and, (b)
is viable.  Most important, utilizing primarily the symmetries of the
theory (rather than detailed dynamics) and the peculiarities of SUSY QCD
as regards forbiddeness of SUSY-breaking, in the absence of gravity,
the preonic
approach provides simple explanations for the desired protection of
composite quark-lepton masses and at the same time for the origins of
family-replication, inter-family mass-hierarchy and diverse mass scales.
It also provides several testable predictions.  In this sense, though
still unconventional, the preonic approach shows promise in being able
to address certain fundamental issues.  I will return to it shortly.
\medskip
(4)~~{\bf Superstrings}: ~~Last but not least, the idea of
superstrings [17] proposes that the elementary entities are not truly
pointlike but are extended stringlike objects with sizes $\sim
(M_{Planck})^{-1}
\sim 10^{-33}$ cm.  Strings with worldsheet supersymmetry, which constitute
superstrings, possess considerable advantage over non-supersymmetric
strings in (a) stabilizing the vacuum and, (b) avoiding large quantum
corrections to Higgs (mass)$^2$ through sub-Planck scale physics.
Furthermore, the superstring theories automatically avoid tachyons.
These theories (which may ultimately be just one) appear
to be most promising in providing a
unified
theory of all the forces of nature including gravity and yielding a
well-behaved
quantum theory of gravity.  In principle, assuming that quarks, leptons
and Higgs bosons are elementary, a suitable superstring theory
could also account for the origin of the three families and the Higgs
bosons at the string unification scale, as well as explain all the
parameters of the standard model.  But in practice, this has not happened as
yet.
Some general stumbling blocks of string theories are associated with
the problems of (i) a choice of the ground state (the vacuum)
from among the many solutions and (ii) understanding supersymmetry breaking.

The ideas listed above are, of course, not mutually exclusive.  In fact the
superstring theories already comprise local supersymmetry and the central
idea of grand unification.  It remains to be seen, however, whether
they give rise, in accord with the standard belief, to elementary
quarks and leptons, or alternatively to a set of substructure fields --
the preons.  In the following, I first
recall the status of conventional grand unification,
and then provide
a perspective as well as motivations for an alternative approach to
grand unification, based on the idea of preons.  In either case, I
highlight the role of supersymmetry.
\bigskip
\noindent
{\rmb III.}~~{\rmb Grand Unification in the Conventional Approach and
Supersymmetry}
\smallskip

By ``Conventional approach'' to grand unification I mean the one in
which quarks and leptons -- and traditionally the Higgs bosons as well
-- are assumed to be elementary [13,14,15].  Within this approach, there
are two distinct routes to higher unification:  (i) the SU(4)-color
route [13] and (ii) SU(5)[14].
Insisting on a compelling reason
for charge -- quantization, the former naturally introduces the
left-right symmetric gauge structure $G_{224}=SU(2)_L \times
SU(2)_R\times SU(4)_{L+R}^C$ [13], which in turn may be embedded in
anomaly-free simple groups like SO(10) or $E_6$ [18].

It has been known for sometime that the dedicated proton decay
searches at the IMB and the Kamiokande detectors [19], and more
recently the precision measurements of the standard model coupling constants
(in
particular ${\rm sin}^2 \hat\theta_W$) at LEP [20] put severe constraints on
grand unification models without supersymmetry.  Owing to such
constraints, the non-SUSY minimal SU(5) and, for similar reasons, the
one-step breaking non-SUSY SO(10)-model, as well, are now excluded
beyond a shadow of doubt.

But the idea of the union of the coupling constants $g_1, g_2,$ and
$g_3$ can well materialize in accord with the LEP data, if one invokes
supersymmetry [21,22,23] into minimal SU(5) or SO(10).  See Fig. 1, which
shows the {\it impressive meeting} of the three coupling constants of
the minimal supersymmetric standard model (MSSM) with an assumed
SUSY-threshold around 1 TeV.  Such a model can, of course, be embedded
within a minimal SUSY SU(5) or SO(10) model, which would provide the
rationale for the meeting of the coupling constants at a scale $M_U
\approx 2 \times 10^{16}$ GeV, and for their staying together beyond
that scale.

The fact that the coupling constants meet in the context of these models
is reflected by the excellent agreement of their predicted value of
$[{\rm sin}^2 \hat\theta_W (m_z)_{theory}= .2325\pm .005$ (using
$\alpha_s(m_z)=\cdot 12\pm \cdot 01)$ with that determined at LEP:
$[{\rm sin}^2\hat\theta_W(m_z)]_{expt.}= .2316\pm .0003$.
In SUSY SU(5) or SO(10), dimension 5 operators do in general pose
problems for proton decay.  But the relevant parameters of the
SUSY-space can be arranged
to avoid conflict with experiments [24].  The SUSY-extensions of SU(5) or
SO(10) typically lead to prominent strange particle decay modes, e.g.,
$p\to \bar\nu K^+$ and $n\to \bar\nu K^0$, while a 2-step breaking of
SO(10) via the intermediate symmetry $G_{224}$ can also lead to
prominent $\Delta (B-L)=-2$ decay modes of the nucleon via Higgs
exchanges such as $p\to e^-\pi^+\pi^+$ and $n\to e^-\pi^+$ and even
$n\to e^-e^+\nu_e$, etc. in addition to the canonical $e^+\pi^0$-mode
[25].

It is encouraging that the super-Kamiokande (to be completed in April
1996) is expected to be sensitive to the $e^+\pi^0$ mode up to partial
lifetimes of few $\times 10^{34}$ years, to the $\bar\nu K^+$ and
$\bar\nu K^0$ modes with partial lifetimes $\leq 10^{34}$ years and to
the non-canonical $n\to e^-e^+\nu_e$ and $p\to e^-\pi^+\pi^+$ modes with
partial lifetimes $< 10^{33}$ years.  Thus the super-Kamiokande,
together with other forthcoming
facilities, in particular, ICARUS,
provide a {\it big
ray of hope} that first of all one will be able to probe much deeper
into neutrino physics in the near future and second proton-decay may
even be discovered within the twentieth century.
\medskip
\noindent
{\bf Questioning the Conventional Approach}

Focusing attention on the meeting of the coupling constants (Fig. 1),
the question arises:  To what extent does this meeting  reflect the
``truth'' or
is it somehow deceptive?  There are two reasons why such a question is
in order.
\smallskip
(1)~~ First, the unity of forces reflected by the meeting of the
coupling constants in SUSY SU(5) or SO(10) is truly incomplete,
because it comprises only the gauge forces, but not the
Higgs-exchange forces.  {\it The latter are still governed by many arbitrary
parameters -- i.e., the masses, the quartic and the Yukawa couplings of
the Higgs bosons  -- and are thus ununified.}  Such arbitrariness goes
against the central spirit of grand unification and has been the main
reason in my mind since the 1970's (barring an important caveat due to
the growth of superstring theories in the 1980's, see below) to consider
seriously the possibility that the Higgses as well as the quarks and the
leptons are composite.  Furthermore, neither SUSY
SU(5) nor SUSY SO(10),
by itself, has the scope of explaining the origins of (a) the three
families, (b) inter- and intra-family mass-splittings and (c) the
hierarchical mass-scales:  from $M_{Planck}$ to $m_\nu$.
\smallskip
(2)~~ The second reason for questioning the conventional approach is
this:  one might have hoped that one of the two schemes -- i.e.,
the minimal SUSY SU(5) or the SUSY SO(10)-model, or a broken ``grand
unified'' symmetry with relations between its gauge couplings near the
string scale, would emerge from one of
the solutions of the superstring theories [17,26], which
would yield not
only the desired
spectrum of quarks, leptons and Higgs bosons but also just the right
parameters for the Higgs masses as well as their quartic and Yukawa
couplings.  While it seems highly nontrivial that so
many widely varying parameters should come out in just the right way
simply from topological and other constraints of string theories,
it would of course be most
remarkable if that did happen.  {\it But so far it has not}.  There are in
fact a very large number of classically allowed degenerate 4D solutions of the
superstring theories (Calabi-Yau, orbifold and free fermionic, etc.),
although one is not yet able
to choose between them.  Notwithstanding this general difficulty of a
choice, it
is interesting that there are at least some
three-family solutions.  However, not
a single one of these has yielded {\it either} a SUSY SU(5) or an
SO(10)-symmetry, {\it or} a broken ``grand unified'' symmetry involving
direct product of groups, with the desired spectrum
{\it and} Higgs-sector parameters, so as to explain the bizarre pattern of
fermion masses
and mixings of the three families [27].  Note that for a string theory
to yield elementary quarks, leptons and Higgs bosons, either the {\it
entire package} of calculable Higgs-sector parameters, which describe
the masses of
all the fermions and their mixings (subject to perturbative
renormalization), should come out just right, or else the corresponding
solution must be discarded.  This no doubt is a {\it heavy burden}.
For the case of the broken grand unified models, there is
the additional difficulty that the grand unification scale of $2\times
10^{16}~GeV$ obtained from low-energy extrapolation does not match
the string unification scale of about $4\times
10^{17}~GeV$ [28].

Thus, even if a certain superstring theory is the right starting point,
and I believe it is, it is not at all clear, especially in view of the
difficulties mentioned above, that it makes contact with the low-energy
world by yielding elementary quarks, leptons and Higgs bosons.  In this
sense, it seems prudent to keep open the possibility that the meeting of
the coupling constants in the context of conventional grand unification,
which after all corresponds to predicting just one number -- i.e.,
$sin^2\theta_W$ -- correctly, may be fortuitous.  Such a meeting should
at least be
viewed with caution as regards inferring the extent to which it
reflects the ``truth'' because there are in fact alternative ways by
which such a meeting can occur (see discussions below).
\medskip
\noindent
{\rmb IV.}~~{\rmb The Preonic Approach to Unification and Supersymmetry}

This brings me to consider an alternative approach to unification based
on the ideas of preons and local supersymmetry [9,10,11,12].  Although
the general idea of preons is old [30], the particular approach [9-12]
which I am about to present has evolved in the last few years.  It is
still unconventional,  despite its promising features.  Its
lagrangian introduces only six positive and six negative chiral preonic
superfields which define the two flavor and four color attributes of a
quark-lepton family and possess only {\it the minimal gauge
interactions} corresponding to flavor-color and metacolor gauge
symmetries [9].  But the lagrangian {\it is devoid altogether of the
Higgs sector since its superpotential is zero owing to gauge and
non-anomalous R-symmetry}.  Therefore, it is free from all the arbitrary
Higgs-mass, quartic and Yukawa coupling parameters which arise in the
conventional approach to grand unification.  This brings {\it real
economy}.  In fact, the preon model possesses just three (or four) gauge
coupling parameters which are the only parameters of the model and even
these few would merge into one near the Planck scale if there is an
underlying unity of forces as we envisage [29].  By contrast, the
standard model has 19 and conventional SUSY grand unification models
have over 15 parameters.  As mentioned in the introductory chapter, in
addition to economy, the main motivations for pursuing the preonic
approach are that it provides simple explanations for (a) the protection
of the masses of the composite quarks and leptons [10], (b) family
replication [11], (c) inter-family mass-hierarchy $(m_{u,d,e}\ll
m_{c,s,\mu}\ll m_{t,b,\pi})$ [12], and (d) diverse mass-scales [9].  At
the same time, it is viable with respect to observed processes including
flavor-changing neutral current processes (see remarks later) and
oblique electroweak corrections.

Fermion-boson partnership in a SUSY theory,
(i.e. $\psi
\leftrightarrow \varphi$ and $v_{\mu} \leftrightarrow \lambda$ or
$\overline{\lambda}$ etc.), leads to several alternative three-particle
combinations with identical quantum numbers, which can make a left-chiral
$SU(2)_L$-doublet family $q^i_L$ -- e.g. (i)~$\sigma_{\mu \nu} \psi^f_L
\varphi^{c^*}_R v_{\mu \nu}$, (ii)~$\sigma^{\mu \nu} \varphi^f_L
\psi^{c^*}_R
v_{\mu \nu}$, (iii)~$\psi^f_L \psi^{c^*}_R \lambda$ and (iv)~$\varphi^f_L
(\sigma^{\mu} \overline{\lambda}) \partial_{\mu} \varphi^{c^*}_R$.  Here
$f$
and $c$ denote flavor and color quantum numbers.
{\it The plurality of these combinations, which stems
because of SUSY, is in essence the origin of family-replication}.
By constructing composite superfields, Babu, Stremnitzer and I
showed  [11] that at the level of minimum dimensional composite
operators
(somewhat analogous to $qqq$ for QCD) there are just three linearly
independent
chiral families $q^i_{L,R}$, and, in addition, two {\it vector-like
families}
$Q_{L,R}$ and $Q^{\prime}_{L,R}$, which couple vectorially to $W_L$'s and
$W_R$'s respectively.  Each of these composite families with spin-1/2 is,
of course,
accompanied by its scalar superpartner  We thus see that one good answer
to Rabi's famous question:  ``Who ordered that?'', is supersymmetry and
compositeness.

Certain novel features in the dynamics of a class of SUSY QCD theories,
in particular (as mentioned in the introduction) the forbidding of
SUSY-breaking in the absence of gravity [8,10], and symmetries of the
underlying preonic theory, play crucial roles in obtaining the other
desired results -- (a), (c) and (d), mentioned above.  The reader is
referred to the papers in Refs. 9-12 and in particular to a recent
review of the preonic approach in Ref. 31 for details of the two broad
dynamical assumptions and the reasons underlying a derivation of these
results.  One attractive feature of the model, which emerges primarily
through the symmetries of the underlying lagrangian, is that the two
vector-like families $Q_{L,R}$ and $Q_{L,R}^\prime$ (mentioned above)
acquire masses of order 1 TeV, while the three chiral families acquire
their masses primarily through their spontaneously induced mixings with
the two vector-like families.  This feature automatically explains why
the electron family is so light compared to the tau-family and (owing to
additional symmetries) why the masses of the muon-family lie
intermediate between those of the electron and the tau-families.  In
particular, the model explains why $m_e\sim 1~MeV$ while $m_t\approx
100-180~GeV$, i.e., why $(m_e/m_t)\sim 10^{-5}$.

Furthermore, using the values of the standard model gauge couplings
measured at LEP and the spectrum of the preon model above and below the
preon-binding scale $\Lambda_M \sim 10^{11}~GeV$, it is found (see Fig.
2) that the flavor-color gauge symmetry being
$SU(2)_L\times U(1)_R\times SU(4)^c$ near the Planck scale and the
metacolor gauge symmetry being either $SU(5)$ [29] or $SU(6)$ [32], the
gauge couplings do tend to meet near the Planck scale.  This opens up a
novel possibility for grand unification at the preon level and thereby a
possible new route for superstring theories to make connection with the
low-energy world.

Last but not least the preon model leads to some {\it crucial
predictions} which include the existence of the two vector-like families
at the TeV-scale.  [See Refs. 12 and and 31 for a list of predictions.]
These two families can be searched for at the forthcoming LHC, the
$e^-e^+$ next linear collider (in planning) and especially at a future
version of the now-extinct SSC.  Their discovery or non-discovery with
masses up to few TeV will clearly vindicate or exclude the preonic
approach developed in Refs. 9-12.
\bigskip
\noindent
{\rmb V.~~Summary and Concluding Remarks}

In summary, we see that just by uniting bosons and fermions, Bose-Fermi
symmetry ends up in playing an {\it essential role} in every attempt at
higher unification, beyond that of the standard model.
\smallskip
$\bullet$~~~~~First, the conventional approach to grand unification,
with elementary quarks, leptons and Higgs bosons, of course needs
supersymmetry, both for a technical resolution of the gauge hierarchy
problem and also to preserve the idea of the meeting of the gauge
coupling constants in accord with the LEP data (i.e., precision
measurement of $\sin^2\Theta_W$).
\smallskip
$\bullet$~~~~~Second, as we saw the alternative preonic approach to
unification requires supersymmetry in a still more crucial manner.  The
Witten index-theorem, which ensures the protection of supersymmetry and
thereby inhibits the formation of supersymmetry and chiral
symmetry-breaking condensate $\langle\bar\psi\psi\rangle$, in the
absence of gravity, plays an essential role in the preonic approach in
that it explains why composite quarks and leptons are so light compared
to their compositeness scale.  In a non-supersymmetric QCD type of
theory, there would be no reason for an inhibition of the
$\langle\bar\psi\psi\rangle$-condensate.  Utilizing local supersymmetry
to its advantage, the preonic approach furthermore provides simple
explanations for the origins of (i) the three chiral families, (ii)
inter-family mass-hierarchy and (iii) diverse mass-scales, and at the
same time (iv) provides the scope for a meeting of the gauge coupling
constants near the Planck scale.  The crucial prediction of the preonic
approach -- that there must exist two vector-like families in the
TeV-range -- is once again tied to supersymmetry.  It is because of
supersymmetry that both chiral and vector-like families arise as
composites naturally from within the model, and also their masses get
tied to the supersymmetry-breaking scale.

In short, because supersymmetry provides some novel features in the
dynamics relative to non-supersymmetric QCD, such as the protection of
chiral symmetry-breaking condensate $\langle\bar\psi\psi\rangle$, and the
scope for a possible breakdown of parity and global vectorial symmetries
like isospin and preon number, it is clear that it would play an even
more crucial role in making ends meet if quarks and leptons are
composite rather than if they are elementary.
\smallskip
$\bullet$~~~~~Last but not least, the idea that the fundamental entities are
not point-like but are one-dimensional strings of sizes $\sim
10^{-33}~cm$ seems to need Bose-Fermi symmetry first of all to stabilize
the vacuum at the string scale, and, second, to avoid large quantum
corrections to Higgs (mass)$^2$ at long distances.  Furthermore,
supersymmetry automatically helps avoid tachyons in string theories.
Strings combined with supersymmetry give rise to superstrings.
As mentioned before, the superstring theories provide the scope for the
greatest synthesis so far in particle physics in that they seem capable
of unifying {\it all matter} (spins 0, 1/2, 3/2, 2 and higher)
as vibrational modes
of the string and also {\it all their interactions}, which include not
only the gauge forces and gravity but also the apparently non-gauge
Higgs-type Yukawa and quartic couplings, within a single coherent
framework.  {\it The most attractive feature is that the superstring theories
permit no dimensionless parameter at the fundamental level}.  Equally
important, they provide the scope for yielding a good quantum theory of
gravity.

For these reasons, I believe that superstring theories possess many (or
most) of the crucial ingredients of a ``final theory'' -- ``the theory of
everything''.  But I also believe that, as they stand, they do not
constitute the whole of an ultimate theory, because, first and foremost,
in spite of the desirable feature that they constrain the gauge symmetry,
the spectrum and
the S-matrix elements (interactions), they are not generated by an
underlying principle analogous to that of general coordinate or gauge
invariance.  Second, as a practical matter, they do not yet explain why we live
in $3+1$ dimensions, and given the fact that supersymmetry does break in
the real world, they do not explain why the cosmological constant is so
small or zero.  Third, they also do not yet provide a consistent understanding
of (a) supersymmetry breaking and (b) choice of the ground state.
Resolutions of some or all of these latter issues, which may well be
inter-related, would clearly involve an understanding of the
non-perturbative aspects and the symmetries of superstring dynamics.
Recent developments which include the ideas of duality symmetries [33]
and the realization that the strong-coupling limit of certain
superstring theories is equivalent to the weak-coupling limit of certain
other theories [34], permitting the elegant and bold conjecture [35] that
there is just one superstring theory, may evolve into a form so as to
achieve the lofty goal of solving superstring dynamics.  It remains to
be seen, however, as to how much of the resolution of the issues
mentioned above could come ``merely'' from our understanding of the
non-perturbative dynamics of the existing string theories and how much of such
a resolution would involve altogether new ingredients (concepts) at
a fundamental level.

As another practical matter, for reasons mentioned in Sections III and
IV, it is
far from clear that the superstring theories make connections with the
low-energy world by yielding elementary quarks, leptons and Higgs
bosons.  The preonic approach, though unconventional, provides a viable
and attractive alternative to the conventional approach.  It therefore
remains to be seen whether the right superstring theory would yield the
elementary quark-lepton-Higgs system with the entire ``right package''
of Higgs-sector parameters or, instead, the preonic spectrum and the
associated gauge symmetry.  In the latter case, the superstring theory
would, of course, be relieved from yielding the right package of such
Higgs sector-parameters because the Higgs-sector is simply absent in
the preonic theory.

To conclude, our understanding of superstring theories is rather
premature.  It would clearly take some time -- optimistically a decade
but conservatively several decades --for us to understand (and
this may be optimistic) the true nature of superstring theories and to
discover the missing ingredients (if any) in these theories, which
together would help resolve the issues mentioned above.  Meanwhile,
regardless of these developments in the future, Bose-Fermi symmetry has
clearly evolved as a great synthesizing principle.  It is a common
denominator and a central feature in all the attempts at higher
unification mentioned above.  Combined with the idea of strings, it
provides the scope, as exhibited in Fig. 3, for unifying matter, forces
and mass-scales.  {\it As such, it is hard to imagine how nature could
have formulated her laws without the aid of supersymmetry}.  It is a
concept which, I believe, is here to stay, analogous to those of general
coordinate and local gauge invariance.  Fortunately, unlike some other
concepts, the relevance of Bose-Fermi symmetry to particle physics can
be established or falsified, depending upon whether the superpartners are
discovered or found to be absent at the forthcoming LEP200, LHC,
$e^-e^+$ NLC and a future version of the now-extinct SSC.

\bigskip
{\rmb VI.~~ Acknowledgements}

The research described in this talk is supported in part by the NSF
grant number 9421387.  This manuscript was prepared during the author's
visit to the Institute for Advanced Study, Princeton, New Jersey, which
is supported in part by a sabbatical leave grant from the University of
Maryland and in part by a grant by the IAS.  It is a pleasure to thank
Keith Dienes, Alon Faraggi, and especially Edward Witten for several
discussions which were
helpful in preparing this manuscript.  I wish to thank Partha Ghose
and Chanchal Majumdar for arranging an excellent meeting in Calcutta and
for their kind hospitality.  I would also like to thank Valerie Nowak
for her most generous cooperation in typing this manuscript.
\medskip

\vfill\eject

\input epsf
\centerline{\epsfxsize 5.0 truein \epsfbox {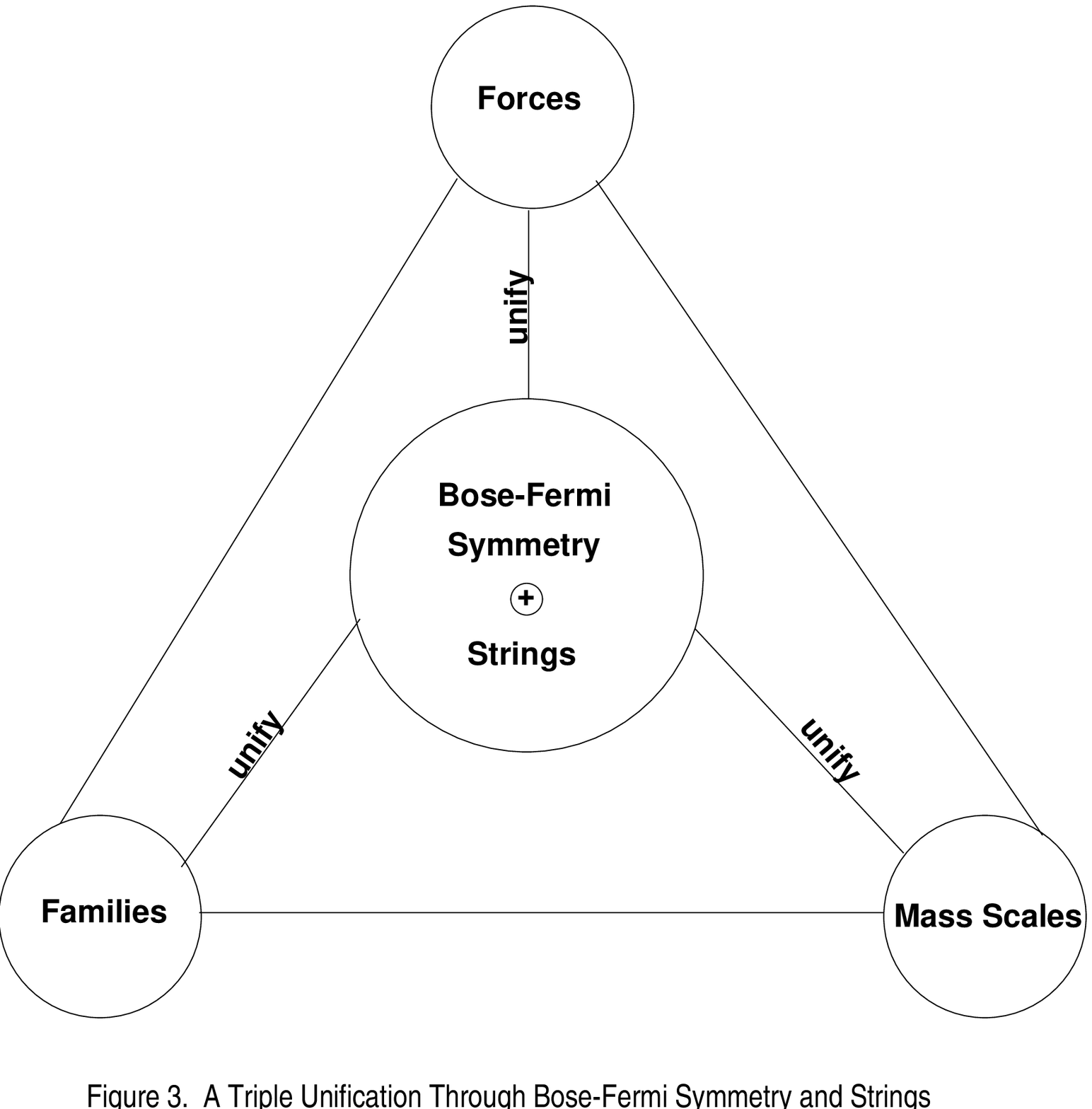}}
\nobreak
 {\bf Figure 3.  A Triple Unification Through Bose-Fermi Symmetry and
     Strings}:
Such a unity of forces, families and mass-scales may arise directly
through a superstring theory, which could give rise to elementary
quarks, leptons and Higgs bosons and fix all the parameters of the
standard model at the string unification scale.  Alternatively, such
a unity may arise through the intermediary of preonic substructures,
which may emerge from superstrings.  These in turn could generate the
three chiral families due to SUSY and a hierarchy of scales owing to
inhibition in SUSY-breaking (see text).  In either case, the twin
ideas of Bose-Fermi symmetry and strings are crucial to achieving
this unification.
\vfill\eject

\centerline{\rmb References}
\item{[1]}
S.N. Bose, letter to A. Einstein, June 4, 1924; Z. Phys. {\bf 26}
(1924) 178.
\item{[2]}
A. Pais, ``Subtle is the Lord,'' Oxford Univ. Press, Page 425.
\item{[3]}
Initiated (for spin-0 particles) by W. Pauli, letter to W. Heisenberg,
June 28, 1934; a review of the different stages of the spin-statistics
theorem may be found in R.F. Streater and A.S. Wightman, ``PCT, Spin and
Statistics and All That,'' Benjamin, New York (1964); and in R. Jost in
``Theoretical Physics in the Twentieth Century,'' eds. M. Fierz and V.
Weisskopf, p. 107, Interscience, New York  (1960).  For a recent review
that includes some new developments, see O.W. Greenberg, D.M. Greenberg
and T.V. Greenbergest in ``Quantum Coherence and Reality,'' eds. J.
Anandan and J.L. Sasko, World Scientific, Singapore (1994), pp. 301-312.
\item{[4]}
Y.A. Gelfand and E.S. Likhtman, JETP Lett. {\bf 13} (1971) 323; J.
Wess and B. Zumino, Nucl. Phys. {\bf B70} (1974)139; Phys. Lett. {\bf
49B}(1974) 52; D. Volkov and V.P. Akulov, JETP Lett. {\bf 16}
(1972) 438.
\item{[5]}
R. Haag, J. Lopuszanski and M. Sohnius, Nucl. Phys. {\bf B88} (1975) 61.
\item{[6]}
S. Coleman and J. Mandula, Phys. Rev. {\bf 159} (1967) 1251.
\item{[7]}
E. Witten, Nucl. Phys. {\bf B188} (1981) 513; R. Kaul, Phys. Lett. {\bf
109B} (1982) 19; S. Dimopoulos and S. Raby, Nucl. Phys. {\bf B192}
(1981) 353.
\item{[8]}
E. Witten, Nucl. Phys. {\bf B185} (1981) 513; {\bf B202} (1983) 253; E.
Cohen and L. Gomez, Phys. Rev. Lett. {\bf 52} (1984) 237.
\item{[9]}
J.C. Pati, Phys. Lett. {\bf B228} (1989) 228.
\item{[10]}
J.C. Pati, M. Cvetic and H. Sharatchandra, Phys. Rev. Lett. {\bf 58}
(1987) 851.
\item{[11]}
K.S. Babu, J.C. Pati and H. Stremnitzer, Phys. Lett. {\bf B256} (1991)
206.
\item{[12]}
K.S. Babu, J.C. Pati and H. Stremnitzer, Phys. Rev. Lett. {\bf 67}
(1991) 1688.
\item{[13]}
J.C. Pati and Abdus Salam; Proc. 15th High Energy Conference, Batavia,
reported by J.D. Bjorken, Vol. 2, p. 301 (1972); Phys. Rev. {\bf 8}
(1973) 1240; Phys. Rev. Lett. {\bf 31} (1973) 661; Phys. Rev. {\bf D10}
(1974) 275.
\item{[14]}
H. Georgi and S.L. Glashow, Phys. Rev. Lett. {\bf 32} (1974) 438.
\item{[15]}
H. Georgi H. Quinn and S. Weinberg, Phys. Rev. Lett. {\bf 33} (1974)
451.
\item{[16]}
For a review see E. Farhi and L. Susskind, Phys. Rev. {\bf 74} (1981)
277 and references therein.
\item{[17]}
M. Green and J. Schwarz, Phys. Lett. {\bf 149B} (1984) 117; D. Gross, J.
Harvey, E. Martinec and R. Rohm, Phys. Rev. Lett {\bf 54} (1985) 502; P.
Candelas, G. Horowitz, A. Strominger and E. Witten, Nucl. Phys. {\bf
B258} (1985) 46.
\item{[18]}
SO(10):  H. Georgi, Proc. AIP Conf. Williamsburg (1994); H. Fritzsch and
P. Minkowski, Ann. Phys. (NY) {\bf 93} (1975) 193.  E(6):  F. Gursey, P.
Ramond and P. Sikivie, Phys. Lett. {\bf 60B} (1976) 177.
\item{[19]}
Particle Data Group, Review of Particle Properties, Phys. Rev. {\bf
D45}, Part II, SI-S584, June 1, 1992.
\item{[20]}
LEP data, Particle Data Group (June, 1994).
\item{[21]}
S. Dimopoulos and H. Georgi, Nucl. Phys. {\bf B193} (1981) 150; N.
Sakai, Z. Phys. {\bf C11} (1982) 153.
\item{[22]}
P. Langacker and M. Luo, Phys. Rev. {\bf D44} (1991) 817; U. Amaldi, W.
de Boer and H. Furstenau, Phys. Lett. {\bf B260} (1991) 447; J. Ellis,
S. Kelley and D.V. Nanopoulos, Phys. Lett. {\bf B260} (1991) 131; F.
Anselmo, L. Cifarelli, A. Peterman and A. Zichichi, Nuov. Cim. {\bf
A104} (1991) 1817.
\item{[23]}
P. Langacker, Review talk at Gatlinburg Conference, June '94,
HEP-PH-9411247.
\item{[24]}
For analysis of this type, see e.g. R. Arnowitt and P. Nath, Phys. Rev.
Lett. {\bf 69} (1992) 725; K. Inoue, M. Kawasaki, M. Yamaguchi and T.
Yanagida, Phys. Rev. {\bf D45} (1992) 328; G.G. Ross and R.G. Roberts,
Nucl. Phys. {\bf B377} (1992) 571; and J.L. Lopez, D. Nanopoulos and H.
Pois, Phys. Rev. Lett. {\bf 47} (1993) 2468.  Other references may be
found in the last paper.
\item{[25]}
J.C. Pati, A. Salam and U. Sarkar, Phys. Lett. {\bf 133B} (1983) 330;
J.C. Pati, Phys. Rev. Rap. Comm. {\bf D29} (1983) 1549.
\item{[26]}
H. Kawai, D. Lewellen and S. Tye, Nucl. Phys. {\bf B288} (1987) 1;
I. Antoniadis, C. Bachas and C. Kounnas, Nucl. Phys. {\bf B289} (1987)
187.
\item{[27]}
Some partially successful three-family solutions with top acquiring a
mass of the right value ($\approx 175~GeV$) and all the other fermions
being massless at the level of cubic Yukawa couplings have been obtained
with $Z_2\times Z_2$ orbifold compactification by A. Farragi, Phys. Lett.
{\bf B274} (1992) 47; Nucl. Phys. {\bf B416} (1994) 63; and J. Lopez, D.
Nanopoulos and A. Zichichi, Texas A\&M preprint CTP AMU-06/95).
In these attempts,
all the other masses and mixings including $m_e\sim {\cal O}(1~MeV)$ are
attributed to in-principle calculable higher dimensional operators.  It
seems optimistic that the entire package of effective parameters
would come out correctly this way with the desired hierarchy.  But, of
course, there is no argument that they cannot.  Thus it seems most
desirable to pursue this approach as far as one can.  This is why I
personally keep
an open mind with regard to both the conventional approach and the
preonic alternative.
\item{[28]}
V.S. Kaplunovsky, Nucl. Phys. {\bf B307} (1988) 145.  Recently,
K.R. Dienes and A.E. Farragi
(preprints hep-th/9505018 and hep-th/9505046)
provide general arguments
why string-threshold corrections arising from the massive tower of
states are naturally suppressed and,
thus, these corrections do not account for such a mismatch between
the two scales.  They and other authors have noted that string theories
tend to give extra matter, which, if they acquire masses in the right
range, could eliminate the mismatch.  For a recent discussion of some
relevant issues pertaining to string-unification, see these papers
as well as L.E. Ib\'a\~nez, talk at Strings '95, USC, March 1995,
FTUAM 95/15-ReV.
\item{[29]}
K.S. Babu and J.C. Pati, Phys. Rev. Rap. Comm. {\bf D48} (1993) R1921.
\item{[30]}
Old works on composite models for quarks {\it and} leptons include the
presently-pursued idea of flavon-chromon preons which was introduced
in the paper of J.C.
Pati and A. Salam, Phys. Rev. {\bf D10} (1974) 275 (Footnote 7).  A
similar idea that treated only quarks but not leptons as composite was
considered independently by O.W. Greenberg (private communication to
JCP).  This idea has been subsequently considered by Pati and Salam in a
set of papers (1975-80) and by several other authors -- with W's treated
as composites in some of them -- see e.g., H. Terezawa, Prog. Theor.
Phys. {\bf 64} (1980) 1763, H. Fritzsch and G. Mandelbaum, Phys. Lett.
{\bf 102B}, (1981) 113, and O.W. Greenberg and J. Sucher, Phys. Lett.
{\bf 99B} (1981) 339; and supersymmetric versions in J.C. Pati
and A. Salam, Nucl. Phys. {\bf B214} (1983) 109; {\it ibid.} {\bf B234}
(1984) 223 and by R. Barbieri, Phys. Lett. {\bf 121B} (1983) 43.  None
of these works provided a reason, however, for (a) the protection of
composite quark-lepton masses, (b) family-replication and (c)
inter-family mass-hierarchy.  The interesting idea of quasi-Nambu
Goldstone fermions suggested by W. B\"uchmuller, R. Peccei and T.
Yanagida, Phys. Lett. {\bf 124B} (1983) 67, provided a partial reason
for some of these issues, in particular for (a), but had problems of
internal consistency as regards SUSY-breaking while maintaining the
lightness of composite quarks and generating effective gauge symmetries.
 The idea of SUSY-compositeness developed in Refs. 9-12 and Ref. 29
introduces a {\it new phase} in the preonic approach in that (i) it avoids the
problems of technicolor and (ii) it seems capable of incorporating the
idea of grand unification [29], while providing a reason for each of the
issues (a), (b) and (c) mentioned above.  It is these features which
seem to make the new approach a {\it viable alternative} to
the conventional approach to grand unification.
\item{[31]}
For a recent review of the preonic approach, see J.C. Pati ``Towards a
Unified Origin of Forces, Families and Mass Scales,''
hep-ph/9505227, to appear in the Proc. of the 1994 Int'l. Conf. on
B-Physics, held at Nagoya, Japan (Oct. 26-29, 1994).
\item{[32]}
The threshold effects at $\Lambda_M$ which permit unity with metacolor
symmetry being SU(6) have been considered by K.S. Babu, J.C. Pati and M.
Parida (to appear).
\item{[33]}
See e.g., A. Sen, Int. J. Mod. Phys. {\bf A9} (1994) 3707, hep-th/9402002
and hep-th/9402032; J.H. Schwarz, hep-th/9411178; C.M. Hull and P.K.
Townsend, QMW94-39, R/94/33 and references therein.
\item{[34]}
E. Witten, preprint IASSNS-HEP-95/18, hep-th/9503124.
\item{[35]}
E. Witten, Ref. 34 and private communications.
\bye